\documentclass[%
reprint,
superscriptaddress,
amsmath,amssymb,
aps,longbibliography,
]{revtex4-2}

\usepackage{graphicx}%
\usepackage{verbatim}
\usepackage[version=4]{mhchem}
\usepackage{xcolor}
\usepackage{physics}
\usepackage[normalem]{ulem}
\usepackage{pdfpages}
\usepackage{pgffor}
\usepackage{booktabs}
\usepackage{ulem}

\usepackage{physics}
\usepackage{xparse}

\newcommand{\mbf}[1]{\ensuremath{\mathbf{#1}}}

\NewDocumentCommand{\rep}{s d<| d|>}{%
\IfBooleanTF{#1}{
   \IfValueTF{#2}{
       \IfValueTF{#3}{\braket{#2}{#3}}{\bra{#2}}
       }{
       \IfValueTF{#3}{\ket{#3}}{}
       }
   }{
   \IfValueTF{#2}{
       \IfValueTF{#3}{\braket*{#2}{#3}}{\bra*{#2}}
       }{
       \IfValueTF{#3}{\ket*{#3}}{}
       }
   }
}

\NewDocumentCommand{\rbra}{sm}{\IfBooleanTF{#1}{\rep<#2|}{\rep*<#2|}}
\NewDocumentCommand{\rket}{sm}{\IfBooleanTF{#1}{\rep|#2>}{\rep*|#2>}}
\NewDocumentCommand{\rbraket}{smom}{
    \IfBooleanTF{#1}{
        \IfNoValueTF{#3}{\rep*<#2||#4>}{\matrixel{#2}{#3}{#4}}
    }{
        \IfNoValueTF{#3}{\rep<#2||#4>}{\matrixel*{#2}{#3}{#4}}
    }
}
\NewDocumentCommand{\field}{o m e{_} e{^} o e{_} e{^}}{
\IfValueTF{#5}{\overline{
  #2\IfValueT{#3}{_#3}\IfValueT{#4}{^{\otimes #4}} %
  \otimes
  #5\IfValueT{#6}{_#6}\IfValueT{#7}{^{\otimes #7}} %
  \IfValueT{#1}{;#1}
}}{
  \IfValueTF{#4}{\overline{
     #2\IfValueT{#3}{_#3}\IfValueT{#4}{^{\otimes #4}}
     \IfValueT{#1}{;#1}
  }}
  {#2\IfValueT{#3}{_#3}}
}
}

\NewDocumentCommand{\frho}{o e{_} e{^}}{
\field[#1]{\rho}_{#2}^{#3}
}
\NewDocumentCommand{\fdelta}{o e{_} e{^}}{
\field[#1]{\delta}_{#2}^{#3}
}

\newcommand{\e}{a}  %

\newcommand{\bx}{\mbf{x}}

\NewDocumentCommand{\ex}{e_}{
\IfValueTF{#1}{\e_{#1}\bx_{#1}}{\e\bx}
}  %

\NewDocumentCommand{\lm}{e_}{
\IfValueTF{#1}{l_{#1}m_{#1}}{lm}
}
\NewDocumentCommand{\nlm}{e_}{
\IfValueTF{#1}{n_{#1}\lm_{#1}}{n\lm}
}
\NewDocumentCommand{\enlm}{e_}{
\IfValueTF{#1}{\e_{#1}\nlm_{#1}}{\e\nlm}
}
\NewDocumentCommand{\en}{e_}{
\IfValueTF{#1}{\e_{#1}n_{#1}}{\e n}
}
\NewDocumentCommand{\nlk}{e_}{
\IfValueTF{#1}{n_{#1}l_{#1}k_{#1}}{nlk}
}
\NewDocumentCommand{\enlk}{e_}{
\IfValueTF{#1}{\e_{#1}\nlk_{#1}}{\e\nlk}
}
\NewDocumentCommand{\enl}{e_}{
\IfValueTF{#1}{\en_{#1}l_#1}{\en l}
}
\NewDocumentCommand{\nl}{e_}{
\IfValueTF{#1}{n_{#1}l_#1}{n l}
}

\NewDocumentCommand{\nnl}{s}{
\IfBooleanTF{#1}{n_1 n_2 l}{n_1; n_2; l}
}
\NewDocumentCommand{\ennl}{s}{
\IfBooleanTF{#1}{\en_1 \en_2 l}{\en_1; \en_2; l}
}

\NewDocumentCommand{\gslm}{s}{
\IfBooleanTF{#1}{\sigma\lambda\mu}{\sigma;\lambda\mu}
}

\newcommand{\prbcorr}[1]{#1} %

\newcommand{\bto}{\ce{BaTiO3}}

\makeatletter
\AtBeginDocument{\let\LS@rot\@undefined}
\makeatother

\begin{document}

\title{Modeling the ferroelectric phase transition in barium titanate \\ with DFT accuracy and converged sampling}

\author{Lorenzo Gigli}
\affiliation{Laboratory of Computational Science and Modeling, Institut des Mat\'eriaux, \'Ecole Polytechnique F\'ed\'erale de Lausanne, 1015 Lausanne, Switzerland}

\author{Alexander Goscinski}
\affiliation{Laboratory of Computational Science and Modeling, Institut des Mat\'eriaux, \'Ecole Polytechnique F\'ed\'erale de Lausanne, 1015 Lausanne, Switzerland}

\author{Michele Ceriotti}
\email{michele.ceriotti@epfl.ch}
\affiliation{Laboratory of Computational Science and Modeling, Institut des Mat\'eriaux, \'Ecole Polytechnique F\'ed\'erale de Lausanne, 1015 Lausanne, Switzerland}
\affiliation{Division of Chemistry and Chemical Engineering, California Institute of Technology, Pasadena, CA, USA}

\author{Gareth A. Tribello}
\email{g.tribello@qub.ac.uk}
\affiliation{Centre for Quantum Materials and Technologies (CQMT), School of Mathematics and Physics, Queen's University Belfast, Belfast, BT7 1NN}

\newcommand{\mc}[1]{{\color{blue}#1}}

\date{\today}%

\begin{abstract}
The accurate description of the structural and thermodynamic properties of ferroelectrics has been one of the most remarkable achievements of Density Functional Theory (DFT). 
However, running large simulation cells with DFT is computationally demanding, while simulations of small cells are often plagued with non-physical effects that are a consequence of the system's finite size. To avoid these finite-size effects one is thus often forced to use empirical models that describe the physics of the material in terms of effective interaction terms, that are fitted using the results from DFT.
In this study we use a machine-learning (ML) potential trained on DFT, in combination with accelerated sampling techniques, to converge the thermodynamic properties of Barium Titanate (BTO) with first-principles accuracy and a full atomistic description.
Our results indicate that the predicted Curie temperature depends strongly on the choice of DFT functional and system size, because of emergent long-range directional correlations in the local dipole fluctuations. 
Our findings demonstrate how the combination of ML models and traditional bottom-up modeling allow one to investigate emergent phenomena with the accuracy of first-principles calculations over the large size and time scales afforded by empirical models. 

\end{abstract}

\maketitle

\section{Introduction}

A ferroelectric is a material that possesses a permanent electric polarization that can be switched under the action of an external electric field. 
Typically, the emergence of the polar phase is accompanied by a structural transition from a high-symmetry paraelectric state down to a broken-symmetry state with a characteristic long-range dipolar ordering \cite{cochran_crystal_1960, jona_shirane_1993}. 
This effect is of wide use in technological applications, for instance in capacitors, whereferroelectrics are used as a high-$\kappa$ dielectric medium for effective energy and charge storage, as well as in piezoelectric devices, sensors and field-effect transistors \cite{muralt_ferroelectric_2000, scott_applications_2007, zhang_lead-free_2021, setter_ferroelectric_2006}. 

Density functional theory (DFT) calculations have complemented experimental efforts that have sought to understand the microscopic structure of ferroelectrics \cite{zhang_ferroelectric_2006,cohen_lattice_1990,ghosez_ab_1998,cohen_origin_1992}. Calculations done at the generalised gradient (GGA) level of theory have revealed details of the phonon instabilities in the cubic high-symmetry structure that cause the $\langle 111 \rangle$ displacements that are predicted by the eight site model~\cite{kotiuga_microscopic_2022,zhao_intrinsic_2022,pasciak_dynamic_2018}. However, studying ferroelectrics using DFT is computationally expensive. In the past researchers have thus been forced to use less-accurate, empirical models to study long time and length scale phenomena. Using such models is no longer necessary as there are now a host of machine learning methods \cite{bartok_gaussian_2010,behl-parr07prl,schutt_schnet_2018,park_accurate_2021,thompson_spectral_2015,shapeev_moment_2016,drautz_atomic_2019,pozdnyakov_smooth_2023,oord_regularised_2020} that offer a way to develop inexpensive models that have DFT-level accuracy for the energies and forces.  There is clearly a problem with using such techniques to study ferroelectrics, however.  Early work \cite{gigli_thermodynamics_2022, xie_ab_2022} in this direction has identified discrepancies between the temperatures at which structural transitions occur in the simulations and the temperatures at which these transitions occur in reality. 
These discrepancies are often reduced by introducing artificial corrections to the simulated pressure \cite{zhong_first-principles_1995, dieguez_ab_2004, tinte_quantitative_2003}. This additional pressure brings the simulation volume closer to the values seen in experiments and drives the predicted Curie temperature towards its experimental value.  
However, this correction tells us little about the physical origins for the discrepancies that are observed in simulations.  

There are multiple sources of error that could be the origin for these discrepancies. (1) The GGA functional (PBEsol) \cite{perdew_restoring_2008} used in previous work slightly underestimates the equilibrium volume of cubic \bto. 
(2) The determination of the transition temperature by tracking spontaneous fluctuations between the phases limits the system size that can be studied, despite the reduced computational cost of the ML potential. 
(3) \prbcorr{Typical machine-learning potentials, such as those based on the SOAP-GAP approach \cite{bart+10prl} or atom-centered symmetry-functions \cite{behl-parr07prl}, rely on descriptors that only account for short-range correlations between the atoms, and are therefore incapable of capturing potentially important long-range, electrostatic interactions. }  

In this work, we systematically investigate the first two of these sources of error.  We fit a new SOAP-GAP potential to energies from the more-accurate, meta-GGA, regularized SCAN functional \cite{furness_accurate_2020}, \prbcorr{as well as the hybrid functional PBE0\cite{perd+96prl,adam-baro99jcp}} and repeat the simulations in our previous work. 
Next, we develop an order parameter that \prbcorr{mimics}  the polarisation of the system. 
This collective variable (CV) allows us to use accelerated sampling to drive the phase transition in larger simulation boxes.  We find that the transition temperature depends strongly on system size, \prbcorr{and trace the substantial finite size effects in this system to the presence of long-ranged dielectric correlations that are similar those observed for effective Hamiltonian models\cite{goncalves_finite_2017}.}

The remainder of this paper is laid out as follows.  We first discuss the calculations that were performed using the regularized SCAN functional in section \ref{sec:ML-SCAN}.  We then describe how we can use atom-centered descriptors to estimate the polarisation in section \ref{sec:CV-ACDC}. Metadynamics simulations that use the polarisation as a CV are then performed in section \ref{sec:metad}. The results from these simulations are accompanied by a discussion of the finite-size effects, which are caused by dipole-dipole correlations.

\section{Methods}

\subsection{Dipole rotation energy barrier and effect of the functional}

As discussed in the introduction, our previous simulations \cite{gigli_thermodynamics_2022} underestimated the temperature at which the system transitions from the tetragonal structure to the cubic one. It may be possible to reduce this discrepancy by employing a more accurate (and expensive) functional for fitting. Before embarking on the fitting we calculated the dipole rotation barrier discussed in \cite{gigli_thermodynamics_2022}, with DFT using VASP \cite{VASP} and the functionals listed in table \ref{table:dipole-rot}. Computational details are provided in the supplemental material. For each functional two single point DFT energy calculations were performed on a distorted version of the rhombohedral (spacegroup R3m) ground state of the system with a 2$\times$2$\times$2 supercell and a lattice parameter of 8~\AA.  In the first of these calculations all the Ti atoms in the system were displaced along the $\langle$111$\rangle$ direction by 0.082~\AA, resulting in aligned local dipoles. In this structure,  Ba and Ti atoms occupy the 1a position ($z_{\text{Ba}} = -0.0004$ and $z_{\text{Ti}} = 0.51116$), while the oxygen occupies the 3b position ($x_{\text{O}} = 0.48823$, $z_{\text{O}} = -0.01872$).
In the second calculation, one of the Ti atoms was again displaced by the same amount in the $\langle\bar{1}\bar{1}\bar{1}\rangle$ direction, while the others were kept fixed. Consequently, in this new configuration, one local dipole is anti-parallel to all the others. The dipole rotation barriers in table \ref{table:dipole-rot} give the difference in energy between these two structures.  
In other words, the dipole rotation barriers in table \ref{table:dipole-rot} are the energies required to flip one dipole. 
Even though this highly-idealized energy difference cannot be taken as a quantitative measure of the energy scale for the ferroelectic phase transition, it tells one something about the energetic cost for disrupting ferroelectric order. 
As we shall see, the same trend we observe here, with more accurate density functionals giving a higher value for this barrier, is qualitatively reflected in a corresponding increase in the ferroelectric transition temperature.

We used the regularized SCAN functional ($\text{r}^2\text{SCAN}$) of Ref. \cite{furness_accurate_2020} to construct the training set for the ML model in this work as this functional provides a good compromise between accuracy and computational cost \cite{sun_strongly_2015}. For reference, one single-point calculation at the PBEsol level, for the aforementioned distorted rhombohedral ground state, only required 109 s on a single HPE Cray node with 128 cpus.
In contrast, the same calculation performed at the PBE0 level required 66313 s on 6 nodes. 
Using the $\text{r}^2\text{SCAN}$ functional required 5940 s per calculation on a single node -- considerably more demanding than a GGA calculation, but not as much as a hybrid functional.

\begin{table}
    \centering
\begin{tabular}{ ccc }
 \toprule
  functional & dipole rotation barrier & diff(\%)\\ 
 \midrule
 PBEsol & 943 meV & \\
 SCAN & 969 meV & +2.8\%\\
 $r^2$SCAN & 991 meV & +5.0\% \\
 PBE0 & 1056 meV & +12\% \\
 \bottomrule
\end{tabular}
\caption{Energy difference between a fully polarized \bto~ structure and a highly-idealized structure with a single anti-parallel dipole,  obtained using various density functionals.  The accuracy and computational expense of the functional increases as you move down the table. The PBEsol estimate of the rotation barrier is computed with Quantum Espresso, the SCAN and $\text{r}^2$SCAN estimates with VASP. Finally, the PBE0 estimate of the rotation barrier is computed both with VASP and Quantum Espresso \cite{giannozzi_quantum_2009}. Notably, they give the same result.}
\label{table:dipole-rot}
\end{table}

\subsection{$\text{r}^2\text{SCAN}$ ML potential}
\label{sec:ML-SCAN}

\begin{figure}
    \centering
    \includegraphics[width=\linewidth]{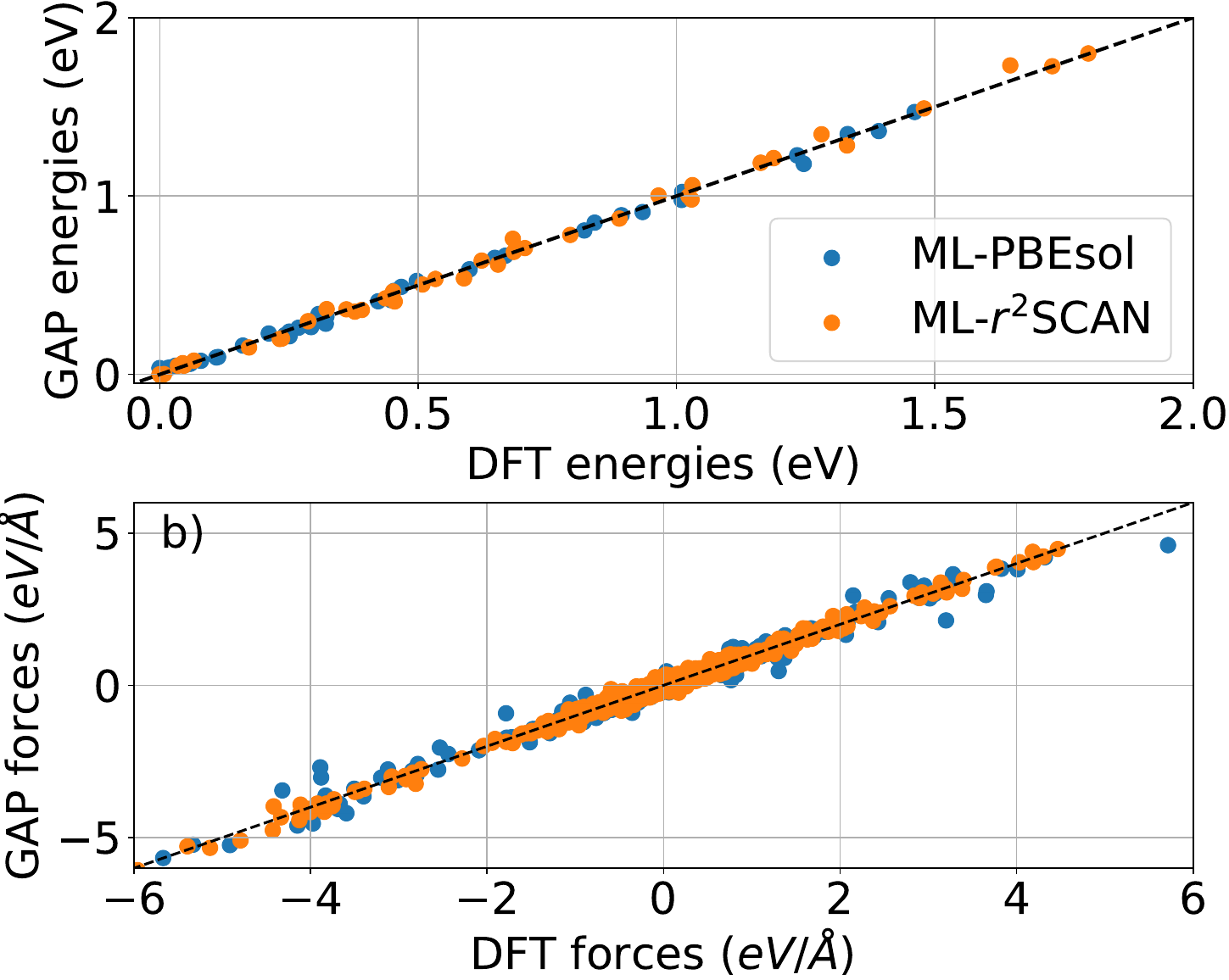}
    \caption{Parity plots for the ML-$\text{r}^2\text{SCAN}$ potential and ML-PBEsol potential of \cite{gigli_thermodynamics_2022}. Panel a) shows the ML energy predictions vs. the DFT energies, while panel b) the force predictions.}
    \label{fig:ML-parity-plots}
\end{figure}

\begin{figure}
    \centering
    \includegraphics[width=\columnwidth]{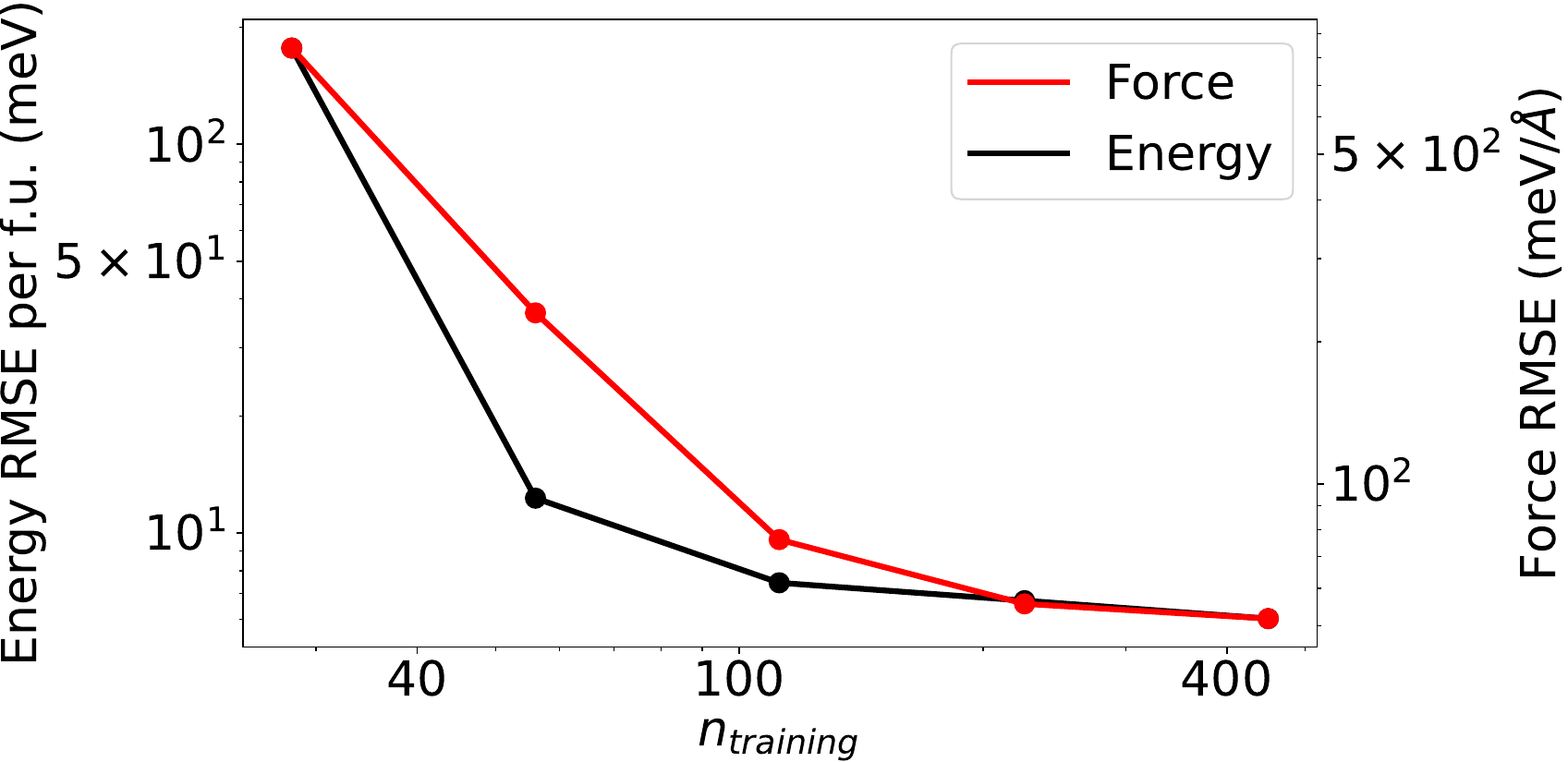}
    \caption{Learning curve for the ML-$\text{r}^2\text{SCAN}$ potential for energy and force predictions. The number of sparse points are kept constant.}
    \label{fig:lc-GAP-model}
\end{figure}

All the ML models discussed in this work were constructed using the SOAP-GAP method \cite{bart+10prl} as implemented in the \texttt{librascal} package \cite{musil_efficient_2021}, using a set of reference energies and atomic forces as target properties. In our previous work, reference energies and forces were extracted from DFT calculations that were performed using the PBEsol functional \cite{gigli_thermodynamics_2022}. This model did not reproduce the experimental critical transition temperatures so for this work we constructed a new model using reference energies and forces calculated with the more accurate $\text{r}^2$SCAN~functional \cite{furness_accurate_2020}. Our original dataset includes structures sampled across the cubic paraelectric and all the ferroelectric phases. Furthermore, the numbers of tetragonal and cubic structures in this data set are comparable. The training set used in this work was built by randomly-selecting $500$ structures from this original dataset and performing single-point DFT calculations using the  $\text{r}^2$SCAN~functional, as implemented in VASP \cite{VASP}.  
We use $450$ structures from this dataset for the fitting procedure and retained $50$ for testing. Parity plots and the learning curve for the resulting ML model predictions on the test set are shown in figures \ref{fig:ML-parity-plots} and \ref{fig:lc-GAP-model}.  
These figures illustrate that the ML validation error for the new model is comparable to that of our old PBEsol model. There is even a slight improvement on the force predictions (see also table \ref{table:RMSEs}). Notably, we obtain this level of accuracy in spite of the fact that only one third of the structures of the original PBEsol dataset were used to train the $\text{r}^2$SCAN model. This dramatic improvement in the efficiency of the training of the SOAP-GAP model is due to the recent implementation of linear system solvers with improved numerical stability using the QR decomposition \cite{numerical_recipes,foster09a} in \texttt{librascal}. 

\prbcorr{ We stress that the final $\text{r}^2\text{SCAN}$ ML model was fitted directly to the $\text{r}^2\text{SCAN}$ DFT data. 
During an initial investigation we found that validation errors were larger when we used $\delta$-learning\cite{Lilienfeld2015} on the PBEsol baseline.  Closer fits were obtained when we fitted the $\text{r}^2\text{SCAN}$ energies and forces directly.}  This result is unusual and may be the result of the different treatments for long range forces in the two functionals. Such differences are difficult to fit with $\delta$-learning as our machine learning model does not use descriptors that describe long-range structural features. 

Details of the DFT calculations and the hyperparameters used in the ML model fitting are provided in the supplemental material.
\prbcorr{As we discuss in the SI, we also fitted a potential to 300 PBE0-level reference calculations. 
Simulations with this potential lead to a large overestimation of the $c/a$ ratio in the tetragonal phase.  This result is consistent with results from previous simulations that were performed with this hybrid functional\cite{evarestov-2012}.
We also observe a dramatic overestimation of the ferroelectric transition temperature, which indicates that r$^2$SCAN is not only more affordable, but also more accurate than PBE0 in predicting the subtle energetic balance that governs ferroelectricity in barium titanate, supporting our choice of this functional as the main focus of this study. 
}

\begin{table}
\begin{tabular}{ cccc }
 \toprule
 RMSE & ML-PBEsol & ML-$r^2$SCAN & ML-PBE0\\ 
 \midrule
 energy (meV/f.u.) & 4.71 & 6.02 & 4.30\\
 forces (meV/$\AA$) & 59.3 (11.1\%) & 51.8 (8.9\%) & 43.0\\
 \bottomrule
\end{tabular}
\label{table:RMSEs}
\caption{RMSEs for energy and force predictions for a 50-structure test set calculated using the ML-PBEsol\cite{gigli_thermodynamics_2022}, the ML-$\text{r}^2$SCAN and the ML-PBE0 model. 1200, 450 and 375 structures, respectively, were used for training. }
\end{table}

\subsection{A collective variable for \bto\,with atom-centered density features}
\label{sec:CV-ACDC}

We use an order parameter based on atom-centered-density features to distinguish between the tetragonal and cubic perovskite structures. In this scheme, the details of which can be found in Refs. \cite{deringer_gaussian_2021, will+19jcp}, a set of neighbor densities around atom $i$ for each chemical element $a$ are calculated using:
\begin{equation}
\rho^{i,a} (\mathbf{r}) = \sum_j \delta_{a, a_j} \exp \left[-\frac{\left|\mathbf{r} - \mathbf{r}_{ji}  \right|^2}{2 \sigma^2}\right] f_{\textrm{cut}}(r_{ji})   
\label{eqn:density}
\end{equation}

In this expression, the sum runs over neighbours $j$ of element $a$ that are within a sphere of radius $r_\textrm{cut}$ centered on atom $i$.  Furthermore, the function $f_\textrm{cut}(r_{ji})$ ensures that the contribution from atom $j$ smoothly goes to zero at $r_\textrm{cut}$.

A quantity that \prbcorr{serves as a proxy for the} local polarisation around atom $i$ is extracted from equation \ref{eqn:density}, by expanding the density using a basis set that is constructed by taking an outer product between a set of Gaussian type orbital (GTO) radial functions and spherical harmonics.  This procedure amounts to calculating the coefficients

\begin{equation}
c_{nlm}^{i,a} = \int \mathrm{d}\mathbf{r}\, R_n(r)^{\ast} Y_l^m(\hat{\mathbf{r}})^{\ast} \rho^{i, a} (\mathbf{r})
\end{equation}
 
The $c_{nlm}^{i,a}$ coefficients for the $l=1$ channel for Ti-centered features offer a representation for the local polarisation as the values of these coefficients change dramatically when Ti atoms are displaced along the 111 direction within the oxygen cage. 
Furthermore, the three $c_{nlm}^{i,a}$ coefficients with $m=+1$, $m=-1$, $m=0$ and $l=1$ (or a sum of these coefficients for the Ti-centered environments) transforms as a vector in real space in the same way as the global polarisation, as motivated in Ref. \cite{gigli_thermodynamics_2022}. 
\prbcorr{As we demonstrate in the SI, these quantities, once summed over all atoms in a structure, correlate very well with the components of the total polarization computed via density-functional perturbation theory.  Using this quantity to represent the polarization is also much cheaper than using the fully-fledged machine-learning model for the polarization that we developed in a previous study\cite{gigli_thermodynamics_2022}. 
We can thus use the components of the density expansion coefficients to  define an effective  local polarisation vector for each atom:}
\begin{equation}
\mathbf{p}^{i}_n = (p^{i}_x, p^{i}_y, p^{i}_z)_n \propto (c^{i,\textrm{O}}_{n1+1}, c^{i,\textrm{O}}_{n1-1}, c^{i,\textrm{O}}_{n10})
\label{eqn:local-pol}
\end{equation}

For each atom in the system there are in principle three (the number of species) times the number of radial basis functions of these 3-dimensional vectors of $c_{nlm}^{i,a}$ components with $l=1$. We use radial basis functions with $n=1$ when calculating the CV and correlations in local polarisations because, as we show in the supplemental material, this CV is better at distinguishing the cubic and tetragonal phases than the symmetrized combination of the cell parameters that we used in our previous work \cite{gigli_thermodynamics_2022}. Therfore, the final CV that was used in this work is given by:  
\begin{equation}
\textrm{P} = \sqrt{ (\textrm{C}_{11-1}^\textrm{O})^2 + (\textrm{C}_{110}^\textrm{O})^2 + (\textrm{C}_{11+1}^\textrm{O})^2 }
\label{eqn:polarisation}
\end{equation}
where:
$$
\textrm{C}_{nlm}^a = \sum_i c_{nlm}^{i,a}
$$
and the sum runs over all central Ti atoms in a given structure. 
\prbcorr{This CV thus corresponds to an approximation of the polarization modulus of a \bto~structure and tracks the structural changes associated with the ferroelectric phase transition, as evidenced in Sec.~\ref{sec:Finite-size-Curie}. }

\subsection{Metadynamics simulations}
\label{sec:metad}

Metadynamics was used to construct a history-dependent bias on the collective variable defined in equation \eqref{eqn:polarisation}. As shown in the supplementary material this bias potential drives transitions between the tetragonal and cubic phases to occur roughly once every 10 ps. This frequency of transition was obtained by using the well-tempered variant of metadynamics \cite{bard+08prl} with $\gamma$ equal to 20 and by adding Gaussian hills with a height of 0.5~kJ mol$^{-1}$ and a width of 0.15 CV-units every 0.1~ps.  \prbcorr{These choices were made based on the fluctuations of the CVs in an unbiased trajectory, as is customary for this field\cite{bussi_using_2020}.}
Simulations were performed using a combination of \texttt{LAMMPS} \cite{plim95jcp}, \texttt{i-PI}, \cite{ipicode, kapil_i-pi_2019}, \texttt{librascal} \cite{musil_efficient_2021} and \texttt{PLUMED} \cite{PLUMED, trib+14cpc}.  Modifications to \texttt{PLUMED} and \texttt{LAMMPS} were required to interface these codes with \texttt{librascal}.  The modified version of these codes and the compilation instructions are available through Refs. \cite{BTO-github, BTO-zenodo}.

The molecular dynamics (MD) simulations were performed in the NST ensemble, with an external diagonal stress tensor $\sigma = \textrm{diag}(\textrm{p}, \textrm{p}, \textrm{p})$ with p = 1 atm. This setup mimics the effect of an isotropic pressure, while keeping the simulation box completely flexible. The cell vectors are all allowed to change, which is essential for ensuring that frequent jumps between the tetragonal and cubic phases can occur. The equations of motion that govern these changes are controlled using a generalized Langevin equation \cite{ceri+11jcp} (GLE). A thermostat thus acts on the cell degrees of freedom.  A second stochastic-velocity-rescaling (SVR) thermostat \cite{buss+07jcp} is then used to control the velocity distribution of the atoms. The characteristic time for the barostat, the SVR thermostat and the MD timestep were set to $1~$ps, $2~$fs and $2~$fs respectively. Simulations with various supercell sizes were performed (from a 4$\times$4$\times$4 cell to a 14$\times$14$\times$14 cell) to achieve finite-size convergence of the relative chemical potentials of the cubic and tetragonal phases.  Furthermore, simulations were performed for temperatures between $150~$K and $320~$K in order to detect the transition point. \texttt{librascal} does not natively support parallelisation over the atoms so we use the domain decomposition implementation in \texttt{LAMMPS} to accelerate the evaluation of the potential. The CV is, however, computed on a single core. To prevent the calculation of the CV from becoming a bottleneck we only computed those features that are necessary to evaluate $\textrm{P}$ and used a multiple time-stepping protocol that computes the bias potential once every $10$ MD steps.  This setup ensures that we are able to generate a nanosecond-long simulation for a system containing 13720 atoms in 5.5 days on one node with 72 CPUs.  Input files and short reference trajectories for our calculations are available from the Materials Cloud \cite{MatCloud}.

\section{Results}

\subsection{Finite-size convergence of the Curie point in \bto}
\label{sec:Finite-size-Curie}

\begin{figure}
    \centering
    \includegraphics[width=\linewidth]{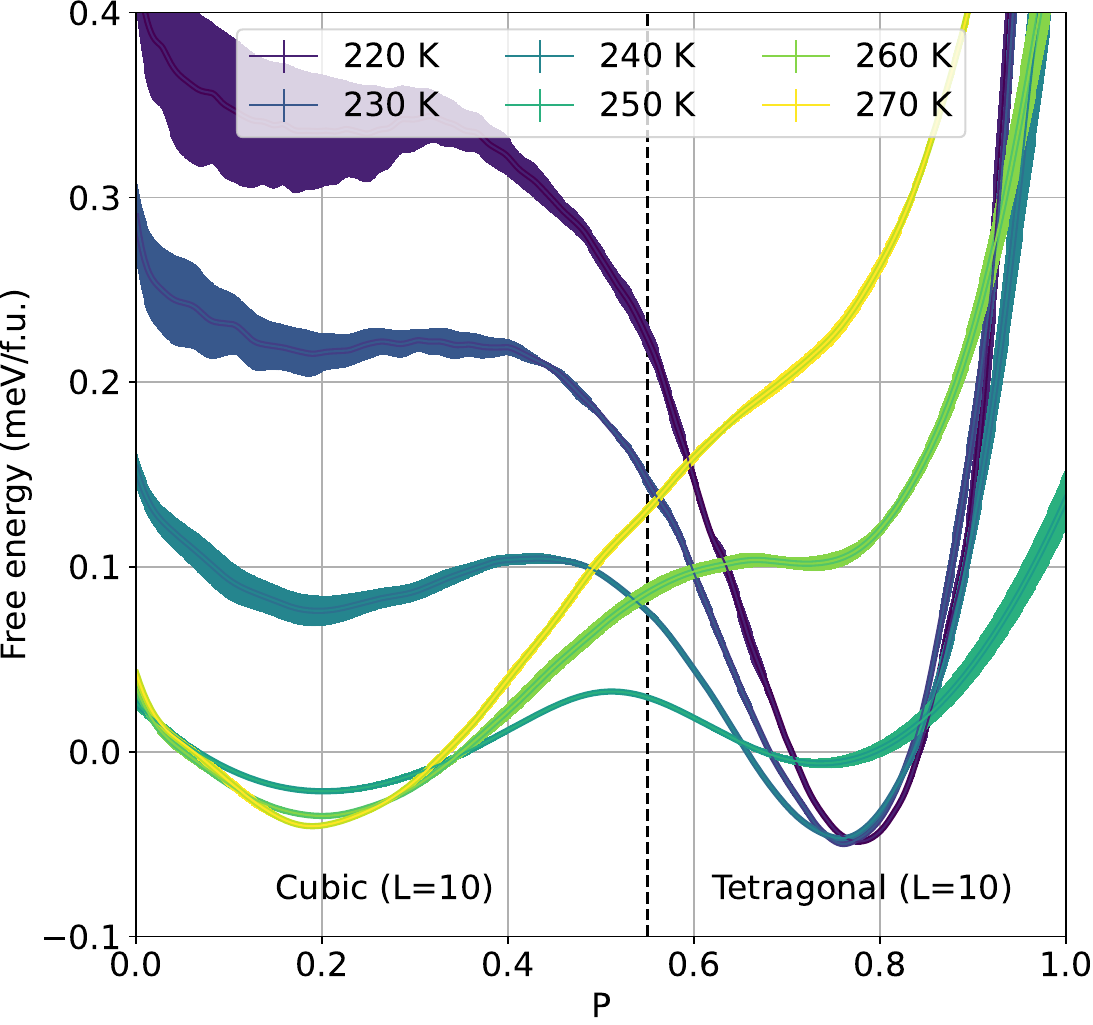}
    \caption{Free energy surfaces as a function of the collective variable for temperatures above and below the transition temperature (249 K) for a $10\times10\times10$ cell. The shaded areas correspond to errors on the free energy, computed as the standard deviation on the mean of the free energy estimates on 4 independent blocks for each metadynamics simulation.}
    \label{fig:1D-free-energy}
\end{figure}

\begin{figure}
    \centering
    \includegraphics[width=\linewidth]{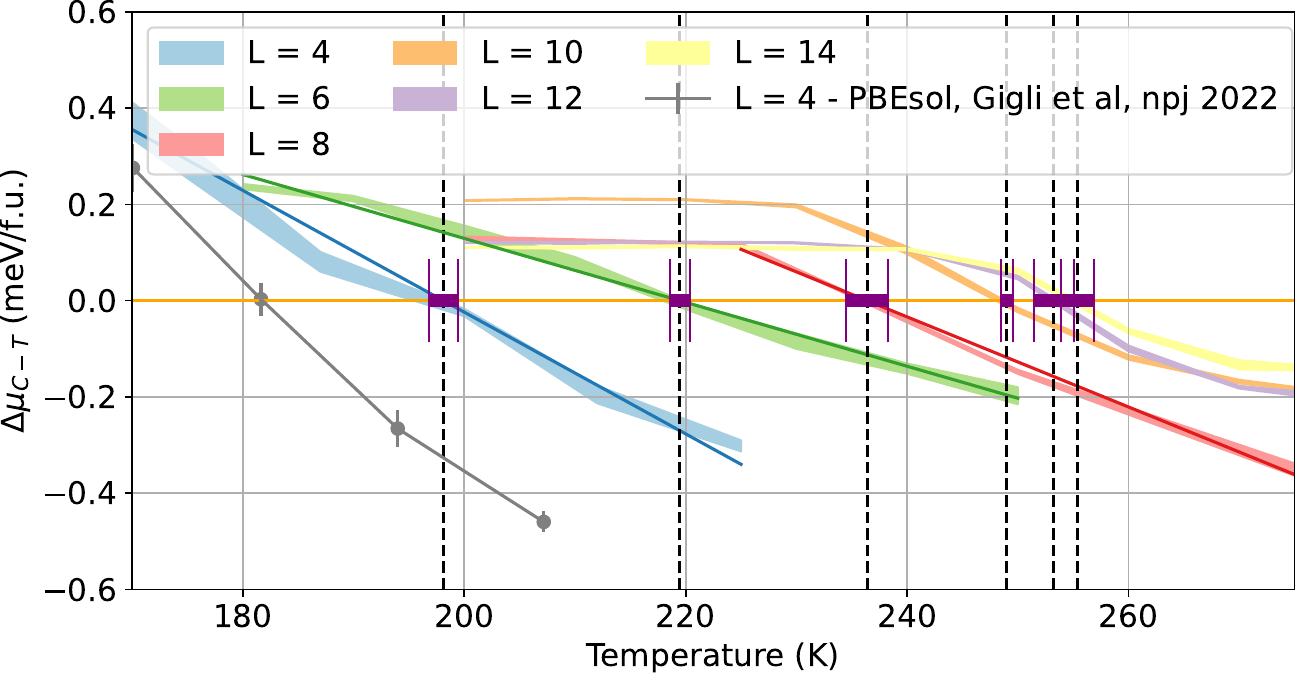}
    \caption{Difference between the chemical potential of the cubic and tetragonal phases as a function of the temperature for different simulation box sizes $L$.}
    \label{fig:size-scaling-delta-mu}
\end{figure}

In our previous work \cite{gigli_thermodynamics_2022} we observed transitions between the cubic and tetragonal phases in unbiased molecular dynamics simulations of $4\times4\times4$ supercells.  These transitions occur in unbiased MD because the cell is relatively small.  When a larger supercell is employed \prbcorr{(and/or a lower temperature is considered, as for the transitions between different ferroelectric phases of \bto)} spontaneous transitions become exceedingly rare and the system remains stuck in the energetic minimum that corresponds to the cubic or tetragonal phase for the duration of the simulation.  
For these systems a simulation bias is thus required to drive transitions between the two phases.  Figure \ref{fig:1D-free-energy} shows that metadynamics simulations using the order parameter described in section \ref{sec:CV-ACDC} can be used to drive  transitions for $10\times10\times10$ supercells. This figure shows the free energy surfaces (FES) that emerge from these metadynamics simulations. These free energy surfaces were obtained by reweighting using the iterative trajectory reweighting (ITRE) method \cite{giberti_iterative_2020}.  Block averaging was used to estimate the errors on the estimates of the free energy shown in figure \ref{fig:1D-free-energy}. 

Figure \ref{fig:1D-free-energy} clearly shows that there is a minimum for high CV values when the temperature is low and the system is in the tetragonal phase and ferroelectric. This minimum is replaced by a minimum at a low value of the CV when the temperature is high and the system is in the cubic phase and paraelectric.  At intermediate temperatures two minima are observed as one phase is metastable.

To extract the difference in chemical potential between the tetragonal and cubic phases we performed clustering using the probabilistic analysis of molecular motifs algorithm (PAMM) \cite{gasp+18jctc}. This clustering technique assigns two probabilities $\theta_1(s_i)$ and $\theta_2(s_i)$ to each CV value $s_i$.  $\theta_1(s_i)$ is the likelihood that the corresponding frame is from the cubic basin, while  $\theta_2(s_i)$ measures the likelihood that it is from is tetragonal basin. The chemical potential difference per formula unit between the two phases can thus be estimated as:
$$
\Delta\mu_{\textrm{C-T}} = -\frac{k_B T}{L^3} \log \left( \frac{\sum_i \theta_1(s_i) w_i}{\sum_i \theta_2(s_i) w_i} \right) 
$$
where the sum runs over all the trajectory frames, $L^3$ is the number of unit cells, $T$ is the temperature and $w_i$ is the weight for each trajectory frame obtained from ITRE.  

\begin{figure}
    \centering
    \includegraphics[width=\linewidth]{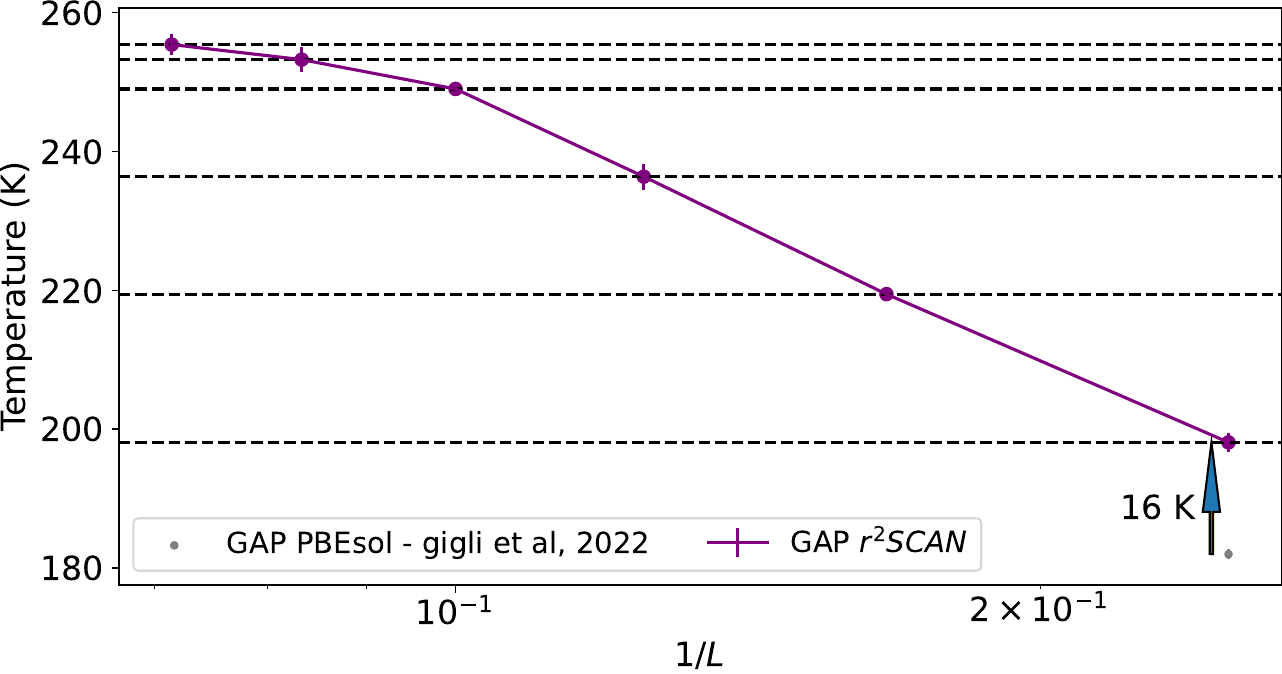}
    \caption{Predicted Curie point at a function of the inverse of the box size $1/L$.  The grey dot at the bottom right shows the result from \cite{gigli_thermodynamics_2022}. The more accurate functional used in this work shifts the transition temperature upwards by 16 K as indicated by the vertical arrow. However, this upward shift is smaller than the increases in transition temperature that are seen for the larger systems simulated in this work.}
    \label{fig:size-scaling-temp}
\end{figure}

Figure \ref{fig:size-scaling-delta-mu} shows how $\Delta\mu_{\textrm{C-T}}$ changes with temperature for a range of differently-sized supercells.  This quantity is initially positive for all cell sizes indicating that the tetragonal phase is more stable than the cubic one at low temperatures.  It becomes negative at high temperatures when the relative stabilities of the two phases reverses.  The Curie point can be determined from figure  \ref{fig:size-scaling-delta-mu} by finding the temperature at which $\Delta\mu_{\textrm{C-T}}$ is zero.  In figure \ref{fig:size-scaling-delta-mu} these temperatures are indicated by the vertical dashed lines.  Errors on these estimates of the Curie temperature are also indicated. To determine these errors we divided each trajectory into four blocks and obtain four separate estimates for each $\Delta\mu_{\textrm{C-T}}$ value.  Variances were computed from these four estimates so the shaded areas in figure \ref{fig:size-scaling-delta-mu} indicate the (1-$\sigma$) confidence limits.  The Curie temperature for each system size was extracted by drawing a line of best fit through the estimates of    $\Delta\mu_{\textrm{C-T}}$.  Propagated errors from this fitting then yield an estimate of the error on the transition temperature.

Figure \ref{fig:size-scaling-delta-mu} clearly shows that the Curie temperature increases with system size.  Furthermore, these differences in transition temperature are for the most part statistically significant.  Figure \ref{fig:size-scaling-temp} indicates the size dependence for the transition temperature more clearly.  In this figure the transition temperature is shown as a function of the inverse box size.  It is only the $14\times14\times14$ system that has a transition temperature that is compatible with the smaller  $12\times12\times12$ system.  For all other system sizes the transition temperature is  underestimated. 
\prbcorr{Furthermore, as shown in the SI, there is also a large simulation-size dependence for  the transition temperature when simulations are performed using the potential that was trained using PBESol.}
The observation of significant finite-size effects here contradicts the analysis provided in Ref. \cite{gigli_thermodynamics_2022} that relied on extrapolating the dielectric constant in the high-temperature regime with a Curie-Weiss law. Interestingly, this indirect approach underestimates the finite size effects. If the aim is to extract accurate thermodynamics simulating large system sizes is thus essential.

\subsection{The role of dielectric correlations}

\begin{figure}
    \centering
    \vspace{-0.3cm}
    \includegraphics[width=\linewidth]{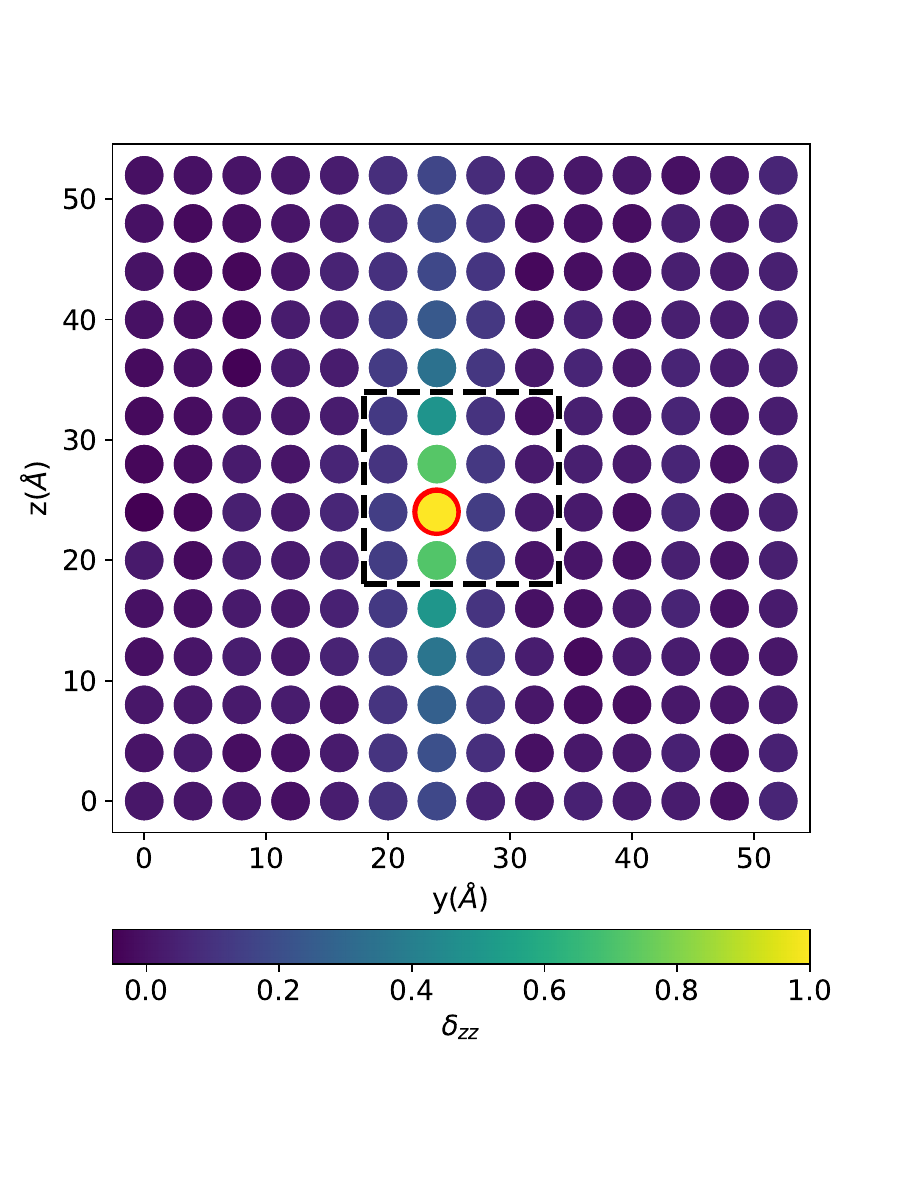}
    \caption{
    Spatial dependence of the $P_{\textrm{z}}$-$P_{\textrm{z}}$ dielectric correlations between atom at the origin (marked by a red circle) and all the other atoms lying on the same (100) plane in a $14\times14\times14$ cell. A dashed rectangle is drawn at the boundary of a $4\times4\times4$ section, corresponding to the smallest cell that we have simulated. This box highlights how much stronger the correlations are at the edge of this small cell than at the boundary of the full cell. }
    \label{fig:rhozz2D}
\end{figure}

\begin{figure}
    \centering
    \includegraphics[width=\linewidth]{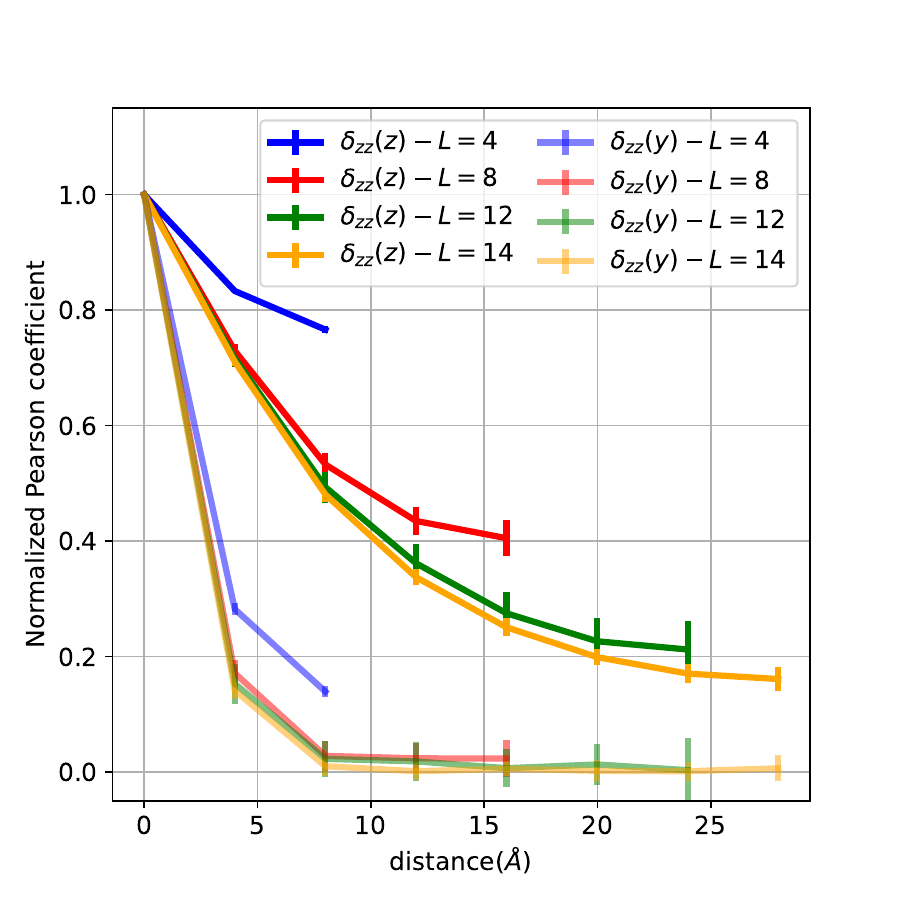}
    \caption{Dielectric correlations in the form of normalized Pearson coefficients, for a set of NST simulations in the cubic regime (above the critical transition temperatures computed in Sec. \ref{sec:Finite-size-Curie}) with different box sizes. The temperatures of these simulations are 225\,K for $L=4$ and 300\,K for $L=8, 12, 14$. The comparison between $\delta_{zz}(z)$ and  $\delta_{zz}(y)$ indicates that correlations for polarizations in the parallel direction are much stronger than in the transverse direction.}
    \label{fig:rhozz}
\end{figure}

We investigated the dielectric correlations to better understand the origin of the enormous system size effects that were described in the previous section.  
To obtain the result shown in figure \ref{fig:rhozz2D} we, therefore, reanalysed the metadynamics simulation on the $14\times14\times14$ cell that was performed at 300 K - a  temperature that is above the transition temperature for the $\textrm{r}^2$SCAN-fitted potential (see also Fig.~\ref{fig:size-scaling-delta-mu}).  
We calculated \prbcorr{a proxy of the}  local polarisation vector for each Ti atom in each frame using equation \eqref{eqn:local-pol} and $n=1$.  We then computed the ensemble average of the following correlation coefficient for each pair of atoms $i$ and $j$ in the system:

\begin{equation}
\label{eq:delta-corr}
\langle \delta_{\alpha\beta}^{ij} \rangle = \left\langle \frac{p_{\alpha}^i p_{\beta}^j}{|\mathbf{p}^i| |\mathbf{p}^j|} \right\rangle
\end{equation}
where $|\mathbf{p^i}|$ is the modulus of $\mathbf{p^i}$ and $\alpha$ and $\beta$ can be x, y or z.  

The coloured circles in figure \ref{fig:rhozz2D} show the average values of $\delta^{0j}_{zz}$ for a (100) slice. The atom at the origin is coloured yellow and marked by a red circle. One can see that most of the other atoms in the system are coloured in purple, which indicates there is relatively little correlation between the polarisation along the z-axis for atom 0 and the polarisation along the z-axis for most of the other atoms. The exceptions to this general rule are the row of atoms that have the same $y$ coordinates as atom 0.  When Ti atoms sit in the same row along $z$ their polarisations prefer to point in the same direction.  This result is consistent with the appearance of the needle-like correlations that are observed both in previous computational studies using DFT and ML models~\cite{akbarzadeh_atomistic_2004, vanderbilt_first-principles_1998,gigli_thermodynamics_2022} and in experiments \cite{bencan_atomic_2021}.

Figure \ref{fig:rhozz} provides a more quantitative assessment of the correlations between local dipoles. In generating this figure we reanalysed the metadynamics simulations on cells of various sizes from the previous section at temperatures where the structure was cubic.  For each box size we computed averages of $\delta_{zz}^{ij}$ over all pairs of atoms that have the same $z$ separation as well as $\delta_{zz}^{ij}$ over all pairs of atoms that have the same $y$ separation. This averaging provides information on the distances over which parallel and transverse correlations between polarisations persist.  One can clearly see that correlations between the $z$-components of the polarisation are much stronger along $z$ than they are along $y$. In other words, the domains tend to polarise along chains that are aligned with the crystallographic directions.  The correlation between these chains is weak and has vanished by the second neighbour. However, even when BTO is in the cubic phase there are no polarisation fluctuations that only affect single Ti atoms. Polarisation fluctuations are highly correlated along the axis of the fluctuation. 
\prbcorr{Similar observations have been made in the literature for \ce{BaTiO3} as well as for other ferroelectrics. The emergence of long-range correlations along the $\langle $100$\rangle$ crystallographic directions was first proposed in Comes \textit{et al} \cite{comes_chain_1968} and theoretically confirmed by Yu and Krakauer in \cite{yu_first-principles_1995} in the case of $\text{KNbO}_3$ by first-principles calculations. 
The latter paper specifically revealed the presence of phonon instabilities associated with chains of displaced Nb atoms. The results presented here are consistent with these findings and justify the use of the CV presented in Sec. \ref{sec:CV-ACDC}, as the polarization proxy defined in Eq. \ref{eqn:local-pol} closely maps the local displacements of Ti-atoms and the local dipoles that are associated with them. 

It is important to reiterate that our ML potential does not consider structural correlations beyond a cutoff of 5.5~\AA{}, as in Ref.~\citenum{gigli_thermodynamics_2022}. The long-range correlations we observe must, therefore,  be generated by an effective short-ranged interaction.  By way of contrast, empirical Hamiltonian models usually employ long-range electrostatic terms. 
} 

\begin{figure}[tp]
    \centering
\prbcorr{
\includegraphics[width=0.9\linewidth]{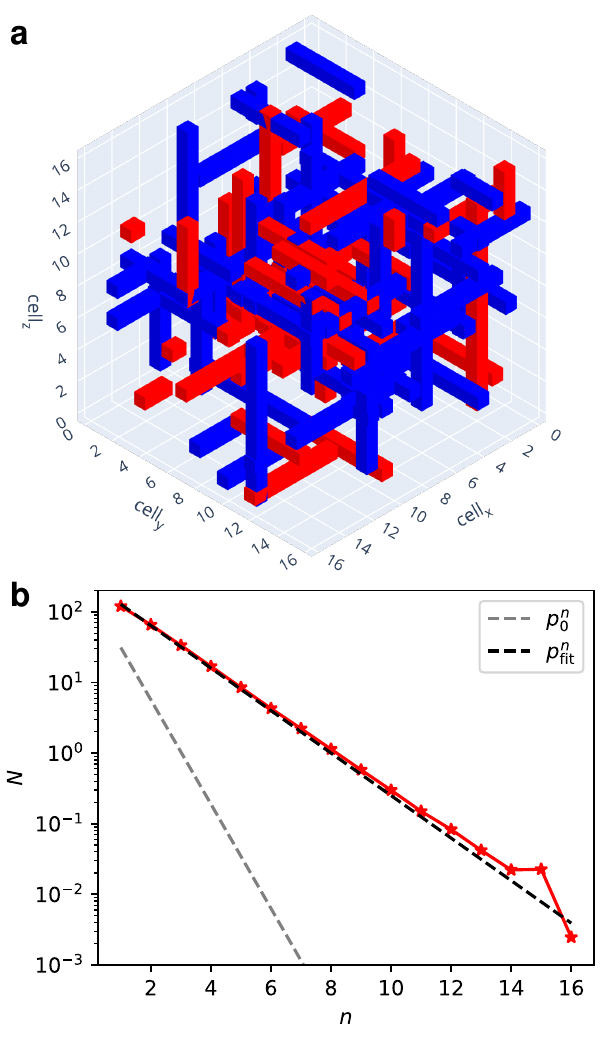}
    \caption{a) Snapshot from a simulation of a $16\times 16\times 16$ cell of \ce{BaTiO3} in the paraelectric phase. Chains of more than 5 adjacent cells with the above-average polarization along one direction are represented with elongated boxes -- red and blue color indicate the direction of the polarization vectors. b) Average counts of polarization chains as a function of length; dashed lines indicate the expected behavior for a random distribution ($p_0$ being the probability of observing a cell with a dipole component above the average), and the dashed black like an exponential decay with a fitted decay rate ($p_{\text{fit}}$).     \label{fig:chain-combo}}
}
\end{figure}

\begin{figure}
    \centering
    \includegraphics[width=\columnwidth]{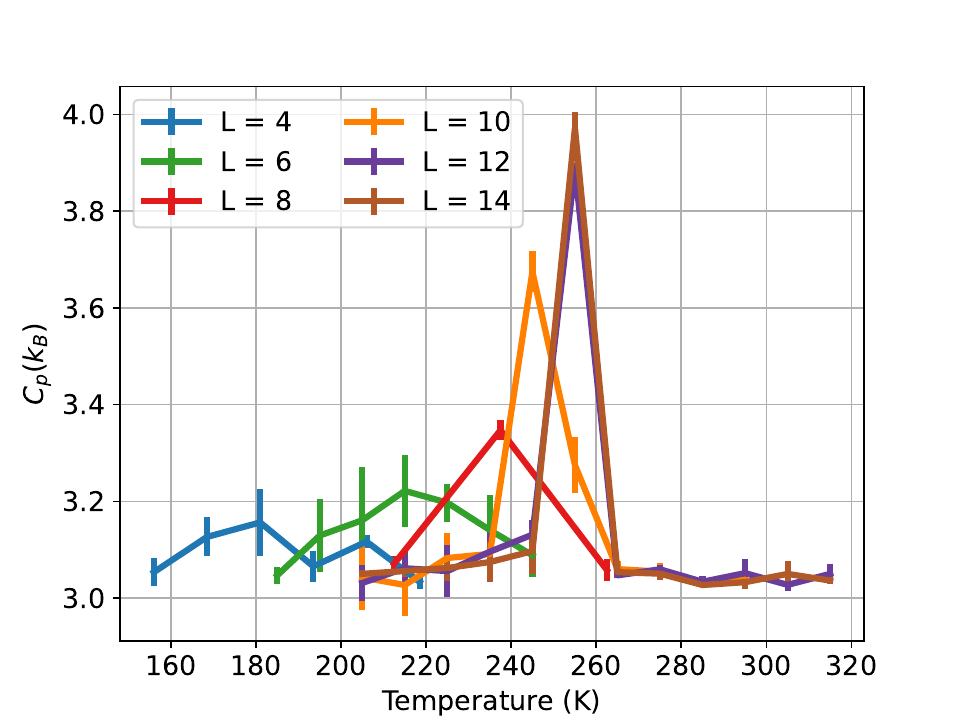}
    \caption{Per-atom heat capacity (expressed in units of $k_{\textrm{B}}$) as a function temperature for NST simulations run with different box sizes. The peak in the heat capacity corresponds to the cubic-tetragonal phase transition. The decrease of the simulation box results in a broadening and a shift of the peak to lower temperatures, consistently with the results of Fig. \ref{fig:size-scaling-temp}.}
    \label{fig:heat-capacity}
\end{figure}

\prbcorr{
In order to better understand the nature of these dipole correlations, we performed an analysis of snapshots from an unbiased simulation at $T=300$~K for a $16\times 16 \times 16$ supercell. 
In this analysis we consider two neighboring cells to be part of a dipole chain along one of the (100) directions if the projection of the dipole on that direction is, for both cells, larger than the root mean square dipole component averaged over the entire trajectory, and aligned in the same direction. 
The procedure yields a ``real-space'' view of the dipole correlations (Fig.~\ref{fig:chain-combo}a) that reveals a seemingly-random coexistence of dipole chains along the three (100) directions, with both polarities.
While the arrangement of the dipole chains is weakly-correlated (consistent with the fast decay of tranverse correlations in Fig.~\ref{fig:rhozz}) there are many more long chains than expected based on a random distribution in the longitudinal direction (Fig.~\ref{fig:chain-combo}b), even though the population decay is still exponential. 
There is even a small but non-zero fraction of chains that span the full size of the supercell, further underscoring the difficulty in converging finite-size effects for this system. 
}

It is possible to use presence of these dipole chains to explain the strong system size effects as these chains will be  affected by the artificial periodicity of the supercell geometry. The square in figure \ref{fig:rhozz2D} shows the extent of the $4\times4\times4$ cell. 
One can clearly see that very-strong correlations extend across the entirety of this small simulation box. In small supercells, the first two coordination spheres around the individual atoms in the cubic phase resemble those around the atoms in the tetragonal phase.  This similarity lowers the energy difference between the two phases, which in turn allows the cubic phase to appear at lower temperatures.
The strength of the correlation for sites on either side of the simulation cell decreases when the cell is larger.  Polarisation sites in larger simulation cells are thus less constrained by their neighbours.  
These sites can thus explore structures that are different from those in the tetragonal phase, which pushes up the energy of the cubic phase.  Even though the cubic phase is higher in energy in these larger cells, it will still form (albeit at higher temperatures) because the greater conformational flexibility ensures that its entropy is higher. 
Figure \ref{fig:heat-capacity} demonstrates that the enthalpy difference of the two phases increases as the box size increases.  
This figure shows the constant pressure heat capacity per atom as a function of temperature in units of $k_B$ for each system size. 
These heat capacities were calculated using finite differences and average enthalpies taken from our metadynamics simulations, as we found that these estimates are better statistically behaved than those obtained from the enthalpy fluctuations. You can clearly see that the heat capacity curve is more strongly peaked when the system size is larger and that the integral of the curve is larger. 
Therefore, these larger peaks for larger system sizes are indicative of a larger per-atom-enthalpy-difference between phases. 
\prbcorr{Similar effects are seen for the lattice expansion (see SI). For a converged supercell (above $10\times 10\times 10$), the temperature dependence of the lattice parameter shows nearly-constant thermal expansion up to the T-C transition temperature, where there is a noticeable discontinuity, consistent with the first-order nature of the phase transition. For smaller supercells, finite-size effects lower the transition temperature and smoothen the discontinuity, altering both the density and its temperature dependence.}

\subsection{2D metadynamics and detection of anisotropic polarizations}
\label{sec:final-results}

\begin{figure*}
    \centering
    \includegraphics[width=\linewidth]{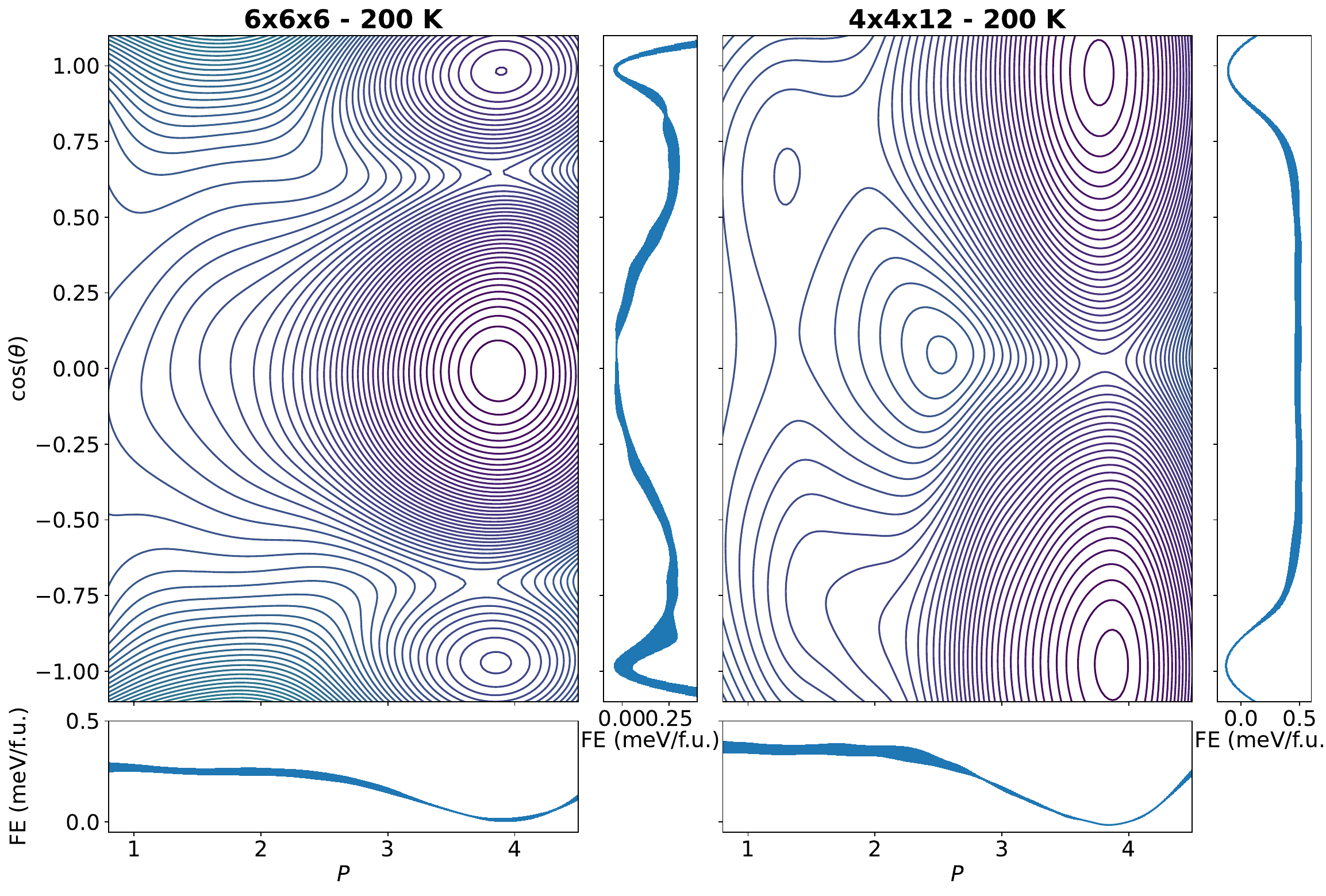}
    \caption{2D free energies for a $6\times6\times6$ cell and $4\times4\times12$ cell at 200 K as a function of the polarization modulus P (as defined in equation~\ref{eqn:polarisation}) and the cosine of the angle ($\cos{\theta}$) that the polarization vector $\mathbf{P}$ forms with the z-axis. Horizontal and vertical sub-panels show the 1-D free energies as a function of P and $\cos{\theta}$, obtained from their respective marginal distributions.}
    \label{fig:2D-metad}
\end{figure*}

Substantial system size effects for the predicted transition temperature is not the only consequence of the long-ranged dielectric correlations that were identified by Figure \ref{fig:rhozz}. This figure indicates that these correlations are also anisotropic. This anisotropy has consequences when one simulates a non-cubic cell as has been recently done in Refs. \cite{kumar_domain_2013, zhou_domain_2015, boddu_molecular_2017, paul_polarization_2008, dimou_pinning_2022,khadka_molecular_2022}. 
We found that problems arise in such simulations because the polarisation for the tetragonal phase can have multiple distinct orientations within the simulation cell. In real systems these distinct orientations are symmetrically equivalent so they should have the same chemical potential. However, we are simulating finite-sized systems.  Consequently, when the simulation box has a non-cubic shape, the orientations that have the polarisation aligned along the long and short axes of the cell are not equivalent. Furthermore, when the polarisation is aligned along the longer axis it is possible to capture more of the long ranged correlations that were identified in Figure \ref{fig:rhozz}.

To investigate whether these anisotropic correlations have a significant effect on the relative energies of phases with different polarisation orientations we introduced the following CV to measure the orientation of the polarisation relative to the lab frame:
$$
\cos(\theta) = \frac{\textrm{C}_{110}}{\textrm{P}}
$$
$\theta$ here measures the the angle between polarisation vector and the $z$-axis for the lab frame.  This CV is thus 1 when the polarisation is parallel to the $z$ axis, -1 when the polarisation  is anti-parallel to the $z$ axis and 0 for the other four orientations that have the polarisation parallel (or anti-parallel) to the $x$/$y$ axes.  

Metadynamics simulations that used $\textrm{P}$ and $\cos(\theta)$ as the CVs were performed for a $6\times 6\times6$ cell and a $4\times 4\times 12$ cell.  The free energies that emerge from these simulations are shown in figure \ref{fig:2D-metad}.  These figures demonstrate that the anisotropy has a significant effect.  The cubic cell has minima in $\cos(\theta)$ at -1, 0 and +1 as would be expected. There is, however, no minimum at 0 for the anisotropic cell.  The tetragonal phase that forms in this simulation always has its polarisation parallel or anti-parallel with the long-axis of the cell.  In other words, formation of tetragonal phases with polarisations aligned along the $x$ and $y$ axis is energetically suppressed.

\section{Conclusions}

Understanding ferroelectricity epitomizes the challenges that are inherent in atomistic modelling of materials. To study this phenomenon one needs accurate models that capture the relationship between the atomic geometry and the electronic structure. 
These models are often computationally expensive, which is problematic as ferroelecticity is an emergent phenomenon that takes place over large length and time scales. It is thus often necessary to use insights that can be extracted from small scale DFT calculations to inform empirical models that can be used to study the long-length and-time-scale behaviours. Machine learning (ML) offers a straightforward method for constructing empirical models from DFT without losing full atomistic detail. In the proceeding sections we have demonstrated the power of this combination by modeling the thermodynamics of the C-T transition in BTO, using a post-GGA level-of-theory in the reference DFT calculations.  Furthermore, we have also shown how ML-inspired order parameters can be used to sample the phase transition in large-scale simulations. 

We find that the transition from the cubic to tetragonal phase occurs at a temperature of 254~K in our simulations.  This value compares much more favourably with the experimental value of 393~K than the value of 182~K that was obtained in previous, similar calculations \cite{gigli_thermodynamics_2022}. Part of the discrepancy can be attributed to the less accurate DFT functional that was used in the previous work.  However, the main source of error comes from very large finite-size effects, that only converge when the simulation box is larger than $12 \times 12 \times 12$ unit cells.  Running such large simulations is impossible with explicit ab initio MD. 

We argue that there are large system size effects in this material because there are long-range directional correlations between local dipoles in BTO. Such correlations have been observed and discussed in the literature \cite{akbarzadeh_atomistic_2004, vanderbilt_first-principles_1998,bencan_atomic_2021} for models that \prbcorr{rely on empirical Hamiltonian models fitted to DFT energetics. 
This paper shows that, when using an unrestricted ML model that yields thermodynamic properties for BTO that are in quantitative agreement with the electronic-structure method used for training\cite{gigli_thermodynamics_2022}, analogous dipole chains with a large longitudinal correlation length are observed. 
This result corroborates the early findings and provides a quantitative assessment for the impact of the details of the DFT calculations.}
\prbcorr{Even though these correlations are usually understood (and modelled) as the consequence of long-range electrostatic interactions, we observe that they also emerge for a ML potential that is restricted, by design, to short-range energetics. 
Investigating the quantitative impact of a model that does not have such limitations (e.g. one based on long-distance equivariants\cite{hugu+23jpcl}) is an interesting future research direction. }
It is also important to note that these 1D dipolar chains can also cause other strong finite-size effects. For example, as we show in section \ref{sec:final-results} when simulating anisotropic cells, the existence of these correlations breaks the symmetry between the different orientations of the tetragonal structure.

This study demonstrates how ML models are an enabling technology for studying thermodynamics and functional properties in materials.  When these methods are used there is no need to  compromise on accuracy, or combine electronic-structure calculations with simplified empirical models. The work presented in this article, serves as a blueprint to tackle similar problems in condensed matter physics and materials science.

\section*{Supplementary materials}

The supplementary materials contain further details on the training procedure, the performance of the different models, and the difficulties in sampling the low-temperature phase transitions. 
Training and validation data, and examples of simulations performed with the various potentials, will be made available upon acceptance.

\begin{acknowledgments}
L.G. and M.C. acknowledge funding from the Swiss National Science Foundation (SNSF) under the Sinergia project CRSII5\_202296 and support from the MARVEL National Centre of Competence in Research (NCCR) for computational resources under the mr31 project (Pillar II). A.G. and M.C. acknowledge funding from the European Research Council Consolidator Grant under the project FIAMMA (588.581).
We would like to acknowledge stimulating discussion with G. Kresse, that triggered a more in-depth investigation of finite-size effects in ML simulations of barium titanate.
\end{acknowledgments}

\end{document}